\documentclass[12pt]{article}
\usepackage{color,epsfig}
\usepackage{graphicx}

\definecolor{violet}{rgb}{0.4,0,0.6}
\definecolor{vert}{rgb}{0,0.4,0.2}
\definecolor{navy}{rgb}{0.0,0.0,0.4}

\def\spose#1{\hbox to 0pt{#1\hss}}\def\lta{\mathrel{\spose{\lower 3pt\hbox
{$\mathchar"218$}}\raise 2.0pt\hbox{$\mathchar"13C$}}}  \def\gta{\mathrel
{\spose{\lower 3pt\hbox{$\mathchar"218$}}\raise 2.0pt\hbox{$\mathchar"13E$}}}

\def\oner{{\color{red}1}} \def\twor{{\color{red}2}}
\def\ivr{{\color{red}i}} \def\jvr{{\color{red}j}}
\def\evr{{\color{red}e}} \def\ebr{{\color{red}{\bf e}}}
\def\fvr{{\color{red}f}} \def\fbr{{\color{red}{\bf f}}}
\def\cvr{{\color{red}c}} \def\cbr{{\color{red}{\bf c}}}
\def\Evr{{\color{red}E}} \def\Ebr{{\color{red}{\bf E}}}
\def\Ivr{{\color{red}I}} \def\Ibr{{\color{red}{\bf I}}}
\def\svr{{\color{red}s}} \def\emptyr{{\color{red}\emptyset}}

   \def\Psiv{{\color{vert}\Psi}}
   \def\Phiv{{\color{vert}\Phi}}

\def\calN{ {\color{red}{\cal N}} }\def\calR{ {\color{red}{\cal R}} }
\def\calP{ {\color{blue}{\cal P}} } \def\PP{{\color{black} P} }
\def\calO{ {\color{blue}{\cal O}} }
\def\Pr{ {\color{vert}{P}} } \def\Pbr{ {\color{vert}{\bf P}} }

\def\be{\begin{equation} } \def\fe{\end{equation}}

\begin{document}

\title{
 \textcolor{red}{\Large Micro-Anthropic Principle for Quantum theory}    
\\[1cm]
}

 \author{Brandon Carter \\
 \textcolor[named]{ForestGreen}{LuTh, Observatoire de Paris - Meudon } 
  }
\date{December, 2004.}
\maketitle

\vskip 1.2 cm

\noindent
{\bf Abstract. } 
Probabilistic models (developped by workers such as Boltzmann, on 
foundations due to pioneers such as Bayes) were commonly regarded merely 
as approximations to a deterministic reality before the roles were 
reversed by the quantum revolution (under the leadership of Heisenberg and  
Dirac) whereby it was the deterministic description that was reduced to 
the status of an approximation, while the role of the observer became 
particularly prominent. The concomitant problem of lack of objectivity
in the original Copenhagen interpretation has not been satisfactorily
resolved in newer approaches of the kind pioneered by Everett. The 
deficiency of such interpretations is attributable to failure to allow 
for the anthropic aspect of the problem, meaning {\it a priori} uncertainty 
about the identity of the observer. The required reconciliation of 
subjectivity with objectivity is achieved here by distinguishing the concept 
of an observer from that of a perceptor, whose chances of identification 
with a particular observer need to be prescribed by a suitable anthropic 
principle. It is proposed that this should be done by an entropy ansatz
according to which the relevant micro-anthropic weighting is taken to be 
proportional to the logarithm of the relevant number of Everett type 
branch-channels.

\vskip 3 cm
{\it Contribution to {\bf Universe or Multiverse?} ed {\rm B.J. Carr}, for 
Cambridge U.P.}
\vfill\eject

\bigskip
\noindent
{\bf 1. Introduction.}
\medskip

As a prescription for ascribing a priori probability weightings to the 
eventuality of finding oneself in the position of particular conceivable 
observers, the anthropic principle was originally developed for application 
to problems of cosmology~\cite{Carter74} and biology~\cite{Carter83}. The 
purpose of the present article is to provide a self contained introductory 
account of the motivation and reasoning underlying the recent 
development~\cite{Carter03} of a more refined version of the anthropic 
principle that is needed for the provision of a coherent interpretation of 
quantum theory.

In order to describe ordinary laboratory applications, it is commonly
convenient, and entirely adequate, to use a ``Copenhagen'' type 
representation in terms of a Hilbert state vector that undergoes 
``collapse''when an observation is made. However from a broader perspective 
it is rather generally recognised that such a collapse can not correspond 
to any actual physical process.

A leading school of thought on this subject was founded by 
Everett~\cite{Everett57}, who maintained the principle of the physical 
reality of the Hilbert state, and deduced that -- in view of the 
agreement that no physical collapse process occurs -- none of the ensuing
branch channels can be ``more real than the rest'', despite the 
paradox posed by the necessity that they be characterised by {\it different} 
(my italics) ``weightings'', of a nature that was never satisfactorily
explained. This intellectual flaw in the Everett doctrine was commonly 
overlooked, not so much by its adherents, who were seriously concerned 
about it~\cite{Graham}, as by its opponents, who were upset 
by its revolutionary ``multi-universe'' implications.

The main alternative line of development was based on the (widely accepted) 
principle -- which will be adopted as the starting point for the present 
work -- that neither the specialised pure Hilbert space vector, nor the 
von Neumann probability operator that replaces it under in more general 
circumstances, is of an objective physical nature, but that they are merely 
mathematical prediction tools of an entirely subjective nature, as also is 
the collapse to which they are subjected if and when the relevant
information becomes available. However this approach also came up against
a paradox, which was exemplified by the parable of 
``Wigner's friend''~\cite{Wigner70} (who, in the more detailed discussion
below, I shall suppose to have been Schroedinger, the owner
of the legendary cat). The problem -- which became particularly
acute in the context of cosmology -- was how independent
observers (such as Wigner and Schroedinger) can be dealt
with objectively, on the same footing, by a probabilistic theory of an
intrinsically subjective nature.

The longstanding  problem of reconciling objectivity with subjectivity
is solved here by the anthropic abstraction, which distinguishes a material
observer (such as Wigner) from that of an abstract perceptor who may
or may not perceive himself to be Wigner. The probability of such
a perception must be attributed by some appropriate micro-anthropic
principle, of the kind~\cite{Carter03} that will be presented below.

\bigskip
\noindent
{\bf 2. Eventualities and observables} 
\medskip

Although their ultimate purpose is to account for (and even
predict) {\it events}, i.e. things that actually happen, physical
(and other) theories are mainly concerned with what I shall refer
to as {\it eventualities}, meaning things that may or not  
actually happen. 

Eventualities are subject to partial ordering, as expressible by a 
statement of the form $\evr_\oner\subset \evr_\twor$, which is to be 
understood as meaning that if an eventuality $\evr_\oner$ happens as an 
actual event then so does $\evr_\twor$. On the understanding that the 
concept of eventuality formally includes the special case of the {\it null} 
eventuality, $\emptyr$, which by definition never happens, it can be taken 
that any pair of eventualities $\evr_\oner$ and $\evr_\twor$ say will 
define a corresponding combined eventuality $\evr_\oner\cap \evr_\twor$ 
whose occurrence as an actual event implies, and is implied by, the 
occurrence, both at once, of $\evr_\oner$ and $\evr_\twor$, so that we 
always have $\evr_\oner\cap \evr_\twor\subset \evr_\oner$. In particular, 
the condition for $\evr_\oner$ to be incompatible with $\evr_\twor$ will be 
expressible as $\evr_\oner\cap \evr_\twor=\emptyr$.

The kinds of (classical and quantum) theory that I know about are all {\it 
additive} in the sense that for each pair of eventualities $\evr_\oner$ and 
$\evr_\twor$ there will be a well defined sum $\evr_\oner\oplus \evr_\twor$ 
that is an admissible eventuality such that $\evr_\oner\cap(\evr_\oner\oplus 
\evr_\twor)=\evr_\oner$. In such a case it is commonly usful to introduce a 
corresponding concept of {\it complementarity} whereby a set $\{ \evr\}$ say 
of  eventualities $\evr_\oner, ..., \evr_\calN$ will be describable as 
complementary in cases for which the sum $\svr=\evr_\oner\oplus ... \oplus 
\evr_\calN$ is an event that must necessarily happen. 

An important related concept -- on which (though it is less fundamental 
than that of an eventuality) discussions of quantum theory are commonly 
based -- is that of an {\it observable}, a  term that is used to describe a 
set $\{\evr\}$ of non-null eventualities that is subject to a condition not 
just of complementarity but also of what may be termed mutual 
{\it exclusivity}. An awkward feature of this concept (one of the reasons 
why I prefer to attribute the primary role to eventualities rather than to 
observables) is that it is difficult to formulate in a manner that 
transcends the technicalities of the particular kind of theory under 
consideration. 

For a theory that is classical, in the sense whose meaning will be 
recapitulated in the next section, a pair of eventualities $\evr_\oner$ and 
$\evr_\twor$ can be considered to be mutually exclusive if they simply 
satisfy the incompatibility condition $\evr_\oner\cap \evr_\twor=\emptyr$, 
but for a quantum theory such incompatibility is merely necessary, but not 
sufficient, for exclusivity in the strong sense -- as defined further on 
below -- that is required for what is meant by observability.

\bigskip

\noindent
{\bf 3. The classical paradigm} 
\medskip

Some of the simplest and most commonly used theories are of the kind 
describable as {\it deterministic}, which means that they consist of rules 
whereby appropriate input data (such as initial conditions) can be use to 
single out a restricted subclass of events that actually happen within a 
much broader class of conceivable eventualities. However a much more widely 
applicable category of theories consists of those that are {\it 
probabilistic}. Instead of providing rules that clearly distinguish events 
that happen from other eventualities that do not, such theories 
merely provide prescriptions for ascribing what is usually called a 
{\it probability} (but what some people prefer to call a {\it propensity}) 
-- meaning a real number $\Pr$ in the range $0\leq \Pr\leq 1$ -- to each of 
the relevant eventualities, in a manner that must naturally be consistent 
with the partial ordering, so that one has $\evr_\oner\supset \evr_\twor\ 
\Rightarrow\ \Pr\{\evr_\oner\}\geq \Pr\{ \evr_\twor\}$ and  in particular 
$\Pr\{\emptyr\}=0$. The category of probabilistic theories evidently 
includes deterministic theories as the special case for which the range of 
probabilities is restricted to the two extreme values, namely $\Pr=1$ 
characterising events, and $\Pr=0$ characterising other eventualities that do 
not actually happen. 

A particularly important subcategory of probabilistic theories is that of 
{\it classical} theories. In a classical theory for the description of a 
system, $A$ say, the admissible eventualities will be identifiable as 
subsets of a corresponding set $\Ivr\{A\}$ that is endowed with an ordinary 
probability measure whose restriction to a subset, $\evr\subset \Ivr$, 
gives the corresponding probability, $\Pr\{\evr\}$, while the complete set 
$\Ivr$ can be interpreted as representing an eventuality that is certain, 
meaning that $\Pr\{\Ivr\}=1$. Such a theory will automatically be endowed 
with an additive structure whereby any pair of eventualities $\evr_\oner$ 
and $\evr_\twor$ will not only have a combination given by the intersection 
$\evr_\oner\cap\evr_\twor$, but it will furthermore have a well defined sum 
that is defined as the corresponding union $\evr_\oner\oplus \evr_\twor=
\evr_\oner\cup \evr_\twor$ so that (unlike what may happen in a quantum 
theory) its probability will be given by $\Pr\{\evr_\oner\oplus \evr_\twor\}
=\Pr\{\evr_\oner\}+\Pr\{\evr_\twor\}-\Pr\{\evr_\oner\cap\evr_\twor\}$.
 
The simplest example of a classical theory applies to the system consisting 
of a tossed coin, which can be described in terms of a total of four 
eventualities. Two of these eventualities are the independent possibilities 
$\evr_\oner$ say for the tail to turn up, and $\evr_\twor$ say for the head 
to turn up, while the other two (trivial) eventualities are their sum 
$\Ivr= \evr_\oner\oplus \evr_\twor$, representing the certain event of 
something turning up, and finally of course the null eventuality $\emptyr
=\evr_\oner\cap\evr_\twor$ representing the impossible case of nothing 
turning up. The latter (trivial) eventualities must always be 
characterised by $\Pr\{\Ivr\}=1$ and $\Pr\{\emptyr\}=0$. The non-trivial 
part of the probability distribution will be given in the unbiased 
version of the theory by $\Pr\{\evr_\oner\}=\Pr\{\evr_\twor\}=\ ^1\!/_2$, 
but could be different in biased versions. In such a (biased or 
unbiased) theory the only non trivial observable consists of the 
complementary pair of alternatives $\{\evr\}=\{\evr_\oner,\evr_\twor\}$, 
but of course there is also the trivial observable consisting just of 
$\Ivr$ by itself.

\bigskip

\noindent
{\bf 4. The Dirac - von Neumann paradigm} 
\medskip

As in a classical theory, the admissible eventualities in a quantum
theory for the description of a system, $A$ say, will be identifiable with 
subsets of a corresponding set $\Ivr\{A\}$. The essential new feature
distinguishing a quantum theory is that $I$ is endowed with a Hilbert 
space structure, and that the {\it admissible eventualities} are 
identifiable, not with arbitrary subsets, but only with those that are 
{\it Hilbert subspaces}.  

If $\evr_\oner$ and $\evr_\twor$ are the Hilbert subspaces representing a 
pair of admissible eventualities, their intersection $\evr_\oner\cap
\evr_\twor$ will also be a Hilbert subspace, representing the corresponding 
conjoint eventuality, but their union $\evr_\oner\cup \evr_\twor$ will in 
general not have the structure of a Hilbert subspace and thus (unlike the 
classical case) will not represent an admissible eventuality. The 
eventualities of a quantum theory do nevertheless have an additive 
structure that is naturally induced by the Hilbert space structure: the sum 
$\evr_\oner\oplus \evr_\twor$ is defined to be the Hilbert subspace that is 
{\it spanned} by the separate Hilbert subspaces $\evr_\oner$ and 
$\evr_\twor$. What this means, using the standard notation scheme 
originally developed by Dirac~\cite{Dirac} (whose lectures on the subject I 
attended as an undergraduate at Cambridge) is that $\vert\Psiv\rangle\ \in 
\ \evr_\oner\oplus \evr_\twor$ if and only if $\vert\Psiv\rangle$ is a 
Hilbert space vector having the form  $\vert\Psiv\rangle=\vert\Psiv_\oner
\rangle+\vert\Psiv_\twor\rangle$ for some pair of Hilbert space vectors 
such that $\vert\Psiv_\oner\rangle\ \in \evr_\oner$ and $\vert\Psiv_\twor
\rangle\ \in \evr_\twor$. In the particular case for which every such pair 
of vectors satisfies the orthogonality condition $\langle \Psiv_\oner\vert
\Psiv_\twor\rangle=0$, the corresponding subspaces$\evr_\oner\subset \Ivr$ 
and $\evr_\twor\subset \Ivr$ will be describable as mutually orthogonal. 

Orthogonality in the sense of the preceding paragraph is what characterises 
the kind of exclusivity required for the definition of what is generally 
known as an observable in the context of quantum theory. Thus an {\it 
observable} (or to be more precise a qualitative observable, as distinct 
from a quantititive observable of the related kind to be discussed below) 
in a quantum theory for the system $A$ can be formally defined to consist 
of a complete set $\{\evr\}$ of mutually orthogonal Hilbert subspaces  
$\evr_\oner$,..., $\evr_\calN$, where the condition of completeness means 
that they span the entire Hilbert space $\Ivr\{A\}$, i.e. that 
$\evr_\oner\oplus ... \oplus \evr_\calN =\Ivr$.

For any particular eventuality, the corresponding subspace $\evr\subset 
\Ivr$ will determine and be determined by an associated Hilbert space 
projection operator ${\bf \evr}= {\ebr}^2$ that is defined -- in such a 
way as to be automatically Hermitean -- by the conditions that ${\ebr}\,
|\Psiv\rangle =\vert\Psiv\rangle$ whenever $\vert\Psiv\rangle$ lies in 
$\evr$, and that ${\ebr}\,\vert\Psiv\rangle =0$ whenever $\vert\Psiv
\rangle$ is orthogonal to the subspace $\evr$. The condition for a set 
$\{\evr\}$ of eventualities, $\{\evr_\ivr\}$ ($\ivr=1, ..., n$) to 
constitute an observable is thus expressible as the condition that the 
corresponding operators should satisfy the orthogonality requirement 
${\ebr}_\ivr\,{\ebr}_\jvr=0$ for $\ivr\neq \jvr$ and that they should 
satisfy the completeness condition $\sum_\ivr {\ebr}_\ivr={\Ibr}$, 
where ${\Ibr}$ is the unit operator on $\Ivr$.

In the earliest versions of quantum theory it was postulated that the 
relevant probabilities would be given just by the specification of a single 
state vector $\vert\Psiv\rangle\ \in \ \Ivr\{A\}$, subject to the 
normalisation condition $\langle\Psiv\vert\Psiv\rangle=1$, according to a 
prescription expressible in the  familiar form 
\be \Pr_{[\calO]}\{\evr_\ivr\}=\langle\Psiv\vert\,\ebr_\ivr\vert\Psiv\rangle
\, .\label{8}\fe
It is to be noted that this is just a conditional probability, subject to
the requirement that the relevant observation, $\calO_\evr$ say, be 
actually carried out.

Soon after the original development of this this Dirac-Heisenberg
paradigm, it came to be recognised that a prescription of the simple 
form (\ref{8}) is too restrictive for applicability to typical cases in 
which the system $A$ under consideration may interact with another 
(internal or external) system, $B$ say. The extended system $\widehat A$ 
say consisting of the combination of $A$ and $B$ will be characterised by 
a Hilbert space $\widehat \Ivr=\Ivr\{\widehat A\}$ that is constructed as 
the tensor product of $\Ivr\{A\}$ and $\Ivr\{B\}$. What this means is that 
a state vector  $\vert\widehat\Psiv\rangle\ \in \ \widehat \Ivr$ for the 
extended system will be expressible in terms of a basis of vectors
$\vert\Phiv_a\rangle\ \in\  \Ivr\{B\}$ satisfying the orthonormality 
condition $\langle\Phiv_a\vert\Phiv_b\rangle=\delta_{ab}$ in the form
\be \vert\widehat\Psiv\rangle=
\sum_{a} |\Phiv_a\rangle|\Psiv_a \rangle\label{9}\fe
for some corresponding set of vectors $\vert\Psiv_a\rangle\ \in \ 
\Ivr\{A\}$ that will not in general be orthonormal, but that must 
satisfy the condition $\sum_a \langle\Psiv_a\vert\Psiv_a\rangle=1$ in 
order for the unit normalisation condition  
$\langle\widehat\Psiv\vert\widehat\Psiv\rangle =1$ to be satisfied. If 
$\evr_\ivr$ is a subspace of dimension $\calR_\ivr$ within the original 
Hilbert space $\Ivr\{A\}$ of dimension $\calN\{A\}$ say, then it will 
determine a corresponding subspace $\widehat \evr_\ivr$ of dimension  
$\calR_\ivr \, \calN\{B\}$ in the tensor product Hilbert space 
$\widehat \Ivr$, where $\calN\{B\}$ is the dimension of $\Ivr\{B\}$. 
Within the original Hilbert space $\Ivr=\Ivr\{A\}$ the corresponding 
projection operator will have rank given by its trace, namely 
$\calR_\ivr={\rm tr}\{{\ebr}_\ivr\}$ while the corresponding operator 
$\widehat {\ebr}_\ivr$ of projection onto $\widehat \evr_\ivr$ in 
$\widehat\Ivr$ will have rank $\calR_\ivr\, \calN\{B\}$.  According 
to the natural extension of the rule (\ref{8}), a unit state vector 
$\vert\widehat\Psiv\rangle$ in $\widehat\Ivr$ will specify a (conditional)
probability distribution given by 
\be \Pr_{[\calO]}\{\evr_\ivr\}=\langle \widehat\Psiv\vert\,
\widehat{\ebr}_\ivr\vert\widehat\Psiv\rangle\, .\label{11}\fe

In order to express such a prescription within the simpler framework
of the original Hilbert space $\Ivr\{A\}$ of the subsystem $A$ with 
which we are particularly concerned, it is necessary to use a prescription
of the kind whose development was attributed by Dirac to von Neumann.

In the Dirac - von Neumann paradigm, instead of being specified just by
a single state vector $\vert\Psiv\rangle$, the (conditional) probability 
distribution (for the outcome of an observation $\calO_\evr$ if
actually performed) is specified by a hermitian probability density 
operator $\Pbr$ say with unit trace ${\rm tr}\{{\Pbr}\}=1$ on $\Ivr$ 
according to the prescription
\be \Pr_{[\calO]}\{\evr_\ivr\}={\rm tr}\{{\Pbr\,\evr}_\ivr\}
\, .\label{14}\fe
This prescription is compatible with the original pure state paradigm, as 
specified just by a single vector satisfying the unit normalisation
$\langle\Psiv\vert\Psiv\rangle=1$, according to the formula (\ref{8})
whose effect can be seen to be the same as that of simply taking
${\Pbr}=\vert\Psiv\rangle\langle\Psiv\vert$ in the general formula 
(\ref{14}). The advantage of the von Neumann type formulation (\ref{14}) is 
that it can also express the result of the more general prescription 
(\ref{11}), whose effect can be seen to be the same as that of taking
\be {\Pbr}=\sum_a\vert\Psiv_a\rangle\langle\Psiv_a\vert\, ,\label{16}\fe
where the (in general non orthonormal) set of vectors 
$\vert\Psi_a\rangle$ is as specified by the decomposition (\ref{9}).

Many authors -- particularly those influenced by the Everett
doctrine~\cite{Everett57} -- have been continued to hanker after the 
original Heisenberg type paradigm -- meaning the supposition that the 
probabilities should ultimately be determined by a pure state in a very 
large all embracing Hilbert space characterising the universe as a whole.
Such authors -- notably including Hawking~\cite{Hawking} -- have been
inclined to regard the use of a von Neumann operator as a rather 
unsatisfactory approximation device that may be made necessary by our 
ignorance due to the regrettable loss of some of the relevant information 
in for example a black hole~\cite{Hawking}. However my own attitude is like 
that of the distrustful insurance agent who doubts whether what was alleged 
to have been lost was ever actually possessed. I personally see no reason 
why -- to encompass more and more detailed microstructure and more and more 
extended macrostructure -- the process of construction of successively  
larger and larger Hilbert spaces should ever come to an end. In other words 
the search~\cite{Vilenkin} for a single ultimate all embracing ``Wave 
function of the universe'', or even of an ultimate all embracing von Neumann 
operator, may be like the pursuit of the proverbially elusive ``Will o' the 
wisp''. It seems more reasonable to accept that any system sufficiently 
simple to be amenable to our mathematical analysis can only be a model of 
an incomplete subcomponent of something larger, and that it is therefore 
unreasonable to demand that it be describable by a pure state rather than a 
more general von Neumann operator. However that may be, these authors would 
agree that there can in any case be no harm in working throughout in terms 
of the von Neumann paradigm, as will be done here, because it includes the 
more restricted Heisenberg type pure state paradigm as a special case.

Before continuing, it is to be remarked that the term  {\it observable}
has been used here to designate what in a more pedantically explicit
terminology would be called a {\it qualitative observable}, in order to
distinguish it from the {\it quantitative observables} that are definable 
as functions thereof.  Thus any qualitative observable, $\{\evr\}$ say, 
determines and is determined by a corresponding equivalence class of 
quantitative variables, in which any particular member, $\Evr$ say, is 
determined by a corresponding non-degenerate real valued function $\Evr_\ivr$ 
of the index labeling the admissible alternatives $\evr_\ivr$ for $\{\evr\}$. 
The condition of non degeneracy of the function is to be understood as 
meaning that $\Evr_\ivr\neq \Evr_j$ whenever $i\neq j$. In a quantum theory 
for a system characterised by a Hilbert space $\Ivr\{A\}$, such a
quantitative observable will be identifiable with a corresponding
Hermitian operator ${\Ebr}$ whose eigenspaces are the Hilbert subspaces 
$\evr_\ivr\ \subset\ \Ivr\{A\}$, while the corresponding eigenvalues are 
the real numbers $\Evr_\ivr$, so that one has 
\be {\Ebr}\vert\Psiv\rangle =\Evr_\ivr\vert\Psiv\rangle\ \Leftrightarrow 
\vert\Psiv\rangle\ \in \ \evr_\ivr\, .\label{10}\fe 
Such a quantitative variable $\Evr$ will have a mean (expectation) value 
$\langle \Evr\rangle$ that will be given by the formula
\be \langle \Evr\rangle ={\rm tr}\{ {\Pbr \Ebr}\}\, ,\label{12}\fe
in which the operator ${\Ebr}$ will be expressible in terms of
the relevant projection operators ${\ebr}_\ivr$  in the explicit form
\be {\Ebr}=\sum_\ivr \Evr_\ivr\, {\ebr}_\ivr\, .\label{13}\fe

The simplest illustration is provided by the familiar Stern Gerlach
example for which the observable ${\Ebr}$ represents the spin energy of 
an electron (with respect to its own rest frame) in a uniform magnetic 
field. For this application, the relevant Hilbert space $\Ivr$ has only two 
(complex) dimensions, being spanned by a subspace $\evr_\oner$ representing 
the eventuality that the spin be aligned with the magnetic field, and a 
subspace $\evr_\twor$ representing the eventuality that it be aligned in the 
opposite direction. Other eventualities, corresponding to alignment in 
other directions, will not be characterised by well defined energy 
values. The quantum analogue of the unbiased coin toss theory considered 
in the previous section is the unbiased spin theory that is specified 
simply by adopting the isotropic probability distribution given (as the
high temperature limit of an ordinary thermal distribution) by 
${\Pbr}=\,^1\!/_2\, {\Ibr}$.

\bigskip
\noindent
{\bf 5. Sensors and conditional probabilities.}
\medskip

Having thus completed a brief overview of the basic quantum mechanical
principles that are generally accepted as a matter of consensus,
it is now necessary to approach the much more controversial issue
of how these rather abstract principles should be interpreted in 
practice -- and more particularly how to relate what might observable 
in principle -- in the abstract sense of the term as used above -- to 
what may be actually observed in the ordinary sense of the word, taking 
it that the ordinary meaning of the word {\it observation}  is the
recognition of the actual occurrence of an eventuality in some
particular system under consideration.

The first, relatively uncontroversial, point that needs to be made at this 
stage is that the notion of an actual observation of an eventuality in a 
generic system under consideration is generally taken to involve an 
interaction with a specialised kind of system that I shall refer to simply 
as a {\it sensor}, which might consist of an artificial measuring apparatus
of a  simple and easily understandable kind such as a Stern Gerlach spin 
orientation detector, but might also consist of something more mysterious 
such as the brain of Schroedinger's famous cat. 

In order for an observable $\{\fvr\}$ say of a system $B$ say under 
consideration to be really able to be (exactly or approximately) observed 
-- i.e. for the recognition of the actual occurrence of a particular 
eventuality $\fvr_j\ \in \{\fvr\}$ to be feasible in practice -- it is 
generally considered to be necessary not just that $\{\fvr\}$ should be 
observable in the abstract sense formulated above, but more particularly 
that it should be adequately correlated with a corresponding sensor 
observable, $\{\evr\}$ say, in an appropriate sensor system, $A$ say. The 
subsets $\widehat \fvr_\jvr=\fvr_\jvr\otimes \Ivr\{A\}$ and  $ \widehat 
\evr_\ivr=\evr_\ivr\otimes \Ivr\{B\}$ in the tensor product space $\widehat 
\Ivr=\Ivr\{A\}\otimes \Ivr\{B\}$ of the combined system will naturally give 
rise to a conjoint observable $\{\cvr\}=\{\evr\}\otimes\{\fvr\}$, whose 
eventualities $\{\cvr_{\ivr\jvr}\}$ are given by the intersection 
subspaces $\widehat \cvr_{\ivr\jvr}=\widehat \evr_\ivr\cap\widehat 
\fvr_\jvr$. The probabilities of these conjoint eventualities will 
evidently form a matrix with elements 
\be \Pr_{\ivr\jvr}=\Pr_{[\calO]}\{\cvr_{\ivr\jvr}\}\, .\label{18}\fe 
The first prerequisite for  the desired correlation of $\{\evr\}$ and 
$\{\fvr\}$ is that they have the same channel number, $\calN_{\!\evr}=
\calN_{\!\fvr}$, i.e. the same number of alternative eventualities, so that 
the matrix $\Pr_{\ivr\jvr}$ will be square. The final requirement for them 
to be more or less adequately correlated is that (for a suitable index 
ordering) the matrix should be more or less exactly diagonal, i.e. that for 
$\ivr\neq \jvr$ the probability $\Pr_{\ivr\jvr}$ should be zero or very 
small. (There is an extensive literature~\cite{Zurek81} on decoherence 
processes by which such diagonalisation can be brought about.)

The conditions of the preceding paragraph are applicable both to classical 
and quantum systems. In the particular case of ordinary quantum systems, 
the observables $\{\evr\}$ and $\{\fvr\}$ will give rise (on the extended 
Hilbert space $\widehat \Ivr$) to corresponding sets of projection operators 
$\widehat{\ebr}_\ivr$ and $\widehat{\fbr}_\ivr$  that will automatically 
commute, $[\widehat {\ebr}_\ivr,\widehat {\fbr}_\jvr]=0$, and whose products 
\be \widehat {\cbr}_{\ivr\jvr}=\widehat{\ebr}_\ivr\widehat{\fbr}_\jvr
=\widehat{\fbr}_\ivr\widehat{\ebr}_\jvr\, \label{19}\fe
will be the projection operators specified by the corresponding subspaces 
$\widehat \evr_\ivr\cap \widehat \fvr_\jvr$ so that (whether it is 
satisfactorily diagonal or not) the probability matrix (\ref{18}) will be 
obtainable from the von Neumann operator $\widehat{\Pbr}$ on $\widehat \Ivr$ 
in the form
\be \Pr_{\ivr\jvr}={\rm tr}\{\widehat{\Pbr}\,\widehat{\ebr}_\ivr
\widehat {\fbr}_\jvr\}\, . \label{20} \fe

It is to be remarked that the relation described in the preceding paragraphs
is {\it reflexive}, in the sense that if an observable $\{\fvr\}$ of $B$ is 
observable by $A$ then the corresponding observable $\{\evr\}$ of $A$ will 
be similarly observable by $B$. A graphic illustration is provided by the 
gedanken experiment in which Schroedinger put his cat in a box that was 
equipped with an anaesthetising mechanism triggered by a Stern Gerlach 
detector. (Schroedinger originally envisaged a lethal mechanism, but that 
would have conflicted with the Popperian desideratum of repeatability of 
the experiment.) One way of describing this is to take the detector to be 
the sensor $A$, whose reading will tell us about the state of the cat, 
considered as system $B$. However by opening the box one can see directly
whether the cat is still awake, thereby using it as a sensor, $A$, that
will tell us whether the spin measured by the detector, now considered as 
system $B$, was up or down. If one also reads the detector as well as 
opening the box, one can check the validity of the theory: an 
inconsistency might remind us of the likelihood for the cat to fall 
asleep spontaneously, with the implication that resort to a less 
satisfactory probability distribution, with non-vanishing off diagonal 
elements, might be more realistic for a subsequent repetition of the
experiment.

It is commonly convenient to rewrite the expression for a joint probability 
such as (\ref{18}) in terms of the corresponding conditional probability 
$\Pr_{[\ivr]}\{\fvr_\jvr\}$ for $\fvr_\jvr$ given $\evr_\ivr$ in the form
\be \Pr_{\ivr\jvr}=\Pr_{[\calO]}\{\evr_\ivr\}\, \Pr_{[\ivr]}\{\fvr_\jvr\}
\, .\label{22} \fe
In the quantum context we are concerned with here, it can be seen that such 
a conditional probablity for $\fvr_\ivr$ will be given by the prescription 
whose form is analogous to that of (\ref{14}), namely
\be \Pr_{[\ivr]}\{\fvr_\jvr\}= {\rm tr}\{ \widehat {\Pbr}_{[\ivr]}
\widehat{\fbr}_\jvr\}\, \label{23} \fe
where $\widehat {\Pbr}_{[\ivr]}$ is the reduced probability operator
associated with the subspace $\widehat \evr_\ivr$, as given in terms of the 
original (unreduced) probability operator $\widehat{\Pbr}$ (on the extended 
space $\widehat \Ivr$) by the defining formula
\be \widehat {\Pbr}_{[\ivr]}= \Pr_\ivr^{\,-1}\widehat{\ebr}_\ivr
\widehat{\Pbr}\,\widehat{\ebr}_\ivr \, .\label{24}\fe
This formula is such as to ensure automatically that the reduced probability
operator has the properties required for qualification as a von Neumann 
density in its own right, meaning that it is Hermitean with unit trace, 
\be{\rm tr} \{\widehat {\Pbr}_{[\ivr]}\}=1\, .\label{25}\fe
The desideratum that $\{\evr\}$ should provide an approximate observation 
of $\{\fvr\}$ is equivalent to the  more restrictive requirement that the 
reduced probability operators should satisfy an approximation of the form
${\rm tr} \{\widehat {\Pbr}_{[\ivr]}\widehat{\fbr}_\jvr\}\approx 
\delta_{\ivr\jvr}$.

\bigskip
\noindent
{\bf 6. The subjective nature of a probability operator.}
\medskip

The consensus about what is meant in quantum theory by a  -- qualitative 
or quantitative -- observable, and by a suitably adapted sensor, in the 
abstract sense does not extend to the question of what is meant by 
the occurrence of an actual observation. There is however a 
rather general understanding that it is something that can be performed 
only by sensors of privileged class for which the title of {\it observer}
is reserved. It would be rather generally agreed, in the context of the 
example referred to above, that this class would include Schrodinger 
himself, but not his (gedanken) Stern Gerlach detector. What is more 
litigious is the status of the cat: would its own discovery that it was 
still awake count as an actual observation?

Such awkward questions are particularly crucial in the context of 
what is commonly referred to as the naive Copenhagen interpretation
(``naive'' to distinguish it from other purportedly more sophisticated
variants) according to which the von Neumann operator - or the state vector 
in the pure case - has  the status of an objective physical entity that 
undergoes a (non unitary) collapse 
\be {\Pbr}\mapsto {\Pbr}_{[\ivr]} \, \label{27} \fe
to the relevant reduced operator (or reduced state vector in the pure case) 
as constructed according to the procedure given by (\ref{24}), when the 
outcome $\evr_\ivr$ is actually observed for an observable $\{\evr\}$. 

The problem with this naive Copenhagen doctrine is how to give a coherent 
prescription for deciding just when this collapse is supposed to occur. 
A relativity theorist would object at the outset that a question about 
when something occurs implicitly refers to the concept of time, a concept 
that is ultimately elusive and at best dependent on a subjectively arbitrary 
choice of reference system. However there also is a more basic problem that 
will arise even in a context for which a reasonably unambiguous Newtonian 
type temporal description is available as a good approximation, as would be 
the case for the cat experiment if not in other (e.g. cosmological) contexts.

This more basic problem~\cite{Wigner70} is that of what is known as 
``Wigner's friend''.  Let us suppose that the friend in question was 
Schrodinger himself, and that Wigner was interested in the fate of the cat.  
Wigner would have had no direct access to the Stern Gerlach detector, but 
would have been able to telephone to Schrodinger to ask what had happened, 
thus using Schroedinger himself as the sensor, which prior to the opening 
of the box would have been in a mixed state. As far as Wigner was concerned 
the relevant collapse process (\ref{27}) would not have been applicable 
until the time of the telephone call, whereas from Schroedinger's point of 
view it would have occurred at the earlier time when the box was opened, 
while the cat itself would have already known even sooner if it had not 
been put to sleep. One might resolve the discrepancy between Schroedinger's 
point of view and that of the cat by taking the line (which might be that of
a theologian such as former Cambridge physics professor, John
Polkinghorne~\cite{Polkinghorne}) that the subhuman status of the cat 
disqualifies it from membership of the privileged class of genuine 
``observers'', but no such specious evasion of the issue is available for 
discrepancy between Wigner's point of view and that of Schroedinger, whose 
equivalent status can not so easily  be denied.

The implication of the well known example recapitulated in the preceding 
paragraph is that the naive Copenhagen interpretation can not be coherently 
applied to cases in which several independent (human or other qualified)
observers are involved, which means that it can ultimately be acceptable 
only to a (deliberate or subconscious) solipsist. 

The obvious conclusion to be drawn from this is that a probability 
operator (or state vector in the pure case) should not be thought of as an 
objective physical entity, and that -- as would be agreed even by
followers of the Everett doctrine, who refuse such subjectivity -- 
its collapse (\ref{27}) should not be thought of as a physical process, 
but just as a mathematical step whose application will be appropriate 
whenever the necessary information, namely the observation of the particular 
eventuality $e_\ivr$, becomes available. The operator collapse process 
(\ref{27}) is thus merely the quantum analogue of the ordinary Bayesian 
reduction process $\Pr\mapsto \Pr_{[\ivr]}$ for an ordinary classical 
probability distribution, whereby its {\it a priori} value is to be replaced 
by the corresponding  {\it a posteriori} -- i.e. conditional -- value when 
the relevant information is supplied. Like the classical probability 
distributions $\Pr$ and $\Pr_{[\ivr]}$, the corresponding {\it a priori} 
and {\it a posteriori} von Neumann operators ${\Pbr}$ and ${\Pbr}_{[\ivr]}$ 
should be considered to have a status that is not objective but 
intrinsically subjective.

A corollary of the foregoing conclusions is the anticipation that
observers with different personal historical backgrounds should use 
{\it different} von Neumann operators, particularly {\it a priori}, 
although there will of course be a tendency toward agreement {\it a 
posteriori} when observational information is shared. In discussions of 
their (different) opinions about what is appropriate in cosmological 
contexts, authors such as Hawking and Vilenkin~\cite{Vilenkin} tend 
to use the definite article for what they call ``the'' state of the 
universe, but the reasoning I am developing here would suggest that such 
definiteness is unjustifiable, and that the most that is reasonable would 
be to propose ``an'' (not ``the'') {\it a priori} probability operator.

\bigskip
\noindent
{\bf 7. Everett's concept of branch-channels.}
\medskip

Having recognised the incoherence of the naive Copenhagen interpretation,
a newer school of thought founded~\cite{Everett57} by Everett has 
emphasised -- correctly according to the reasoning I am developing here 
-- that there is no physical process of collapse of the probability 
operator. What is not so clearly correct or even meaningful is Everett's 
concomitant conclusion that all the ensuing ``branches of the universe'' 
remain equally real. 

Before the validity of this doctrine can be discussed, it is necessary to 
explain what is meant by the branches -- or to be more precise
branch-channels -- in question. The origin of the idea dates back to the 
pre von Neumann epoch when it was assumed that the relevant probability 
distribution would be provided by a pure state, as specified by a unit 
Hilbert space vector that could of course be represented as a sum, 
$\vert\Psiv\rangle=\sum_\ivr\vert\Psiv_\ivr\rangle$, of eigenvectors 
$\vert\Psiv_\ivr\rangle\ \in \ \evr_\ivr$ of the observable $\{\evr\}$ 
under consideration. The observation process was commonly described
as having a first step consisting of a splitting of $\vert\Psiv\rangle$ 
into the set of alternative projections 
\be \vert\Psiv_\ivr\rangle={\ebr}_\ivr\vert\Psiv\rangle\, \label{29}\fe
onto the relevant eigenspaces, which were referred to (rather misleadingly) 
as branches. 

             According to the naive Copenhagen doctrine, the observation 
process would be completed by a second step consisting of a collapse, 
whereby the set would be replaced by a single appropriately renormalised 
branch vector, $\Pr_\ivr^{\,-1/2}\vert\Psiv_\ivr\rangle$ that would turn up 
with the corresponding conditional probability 
$\Pr_\ivr=\langle\Psiv_\ivr\vert\Psiv_\ivr\rangle$. 
On the other hand the Everett doctrine denied the occurrence of the 
collapse as a physical process, with the implication that the system would 
be subsequently describable~\cite{Neumann} as being in a mixed state, for 
which the corresponding von Neumann operator would have the form 
\be {\Pbr}=\sum_\ivr|\Psiv_\ivr\rangle\langle\Psiv_\ivr\vert
\, ,\label{31}\fe
representing what I shall refer to as the {\it provisional} probability 
operator, in order to distinguish it from the relevant (pure) {\it a 
priori} probability operator 
\be {\Pbr}_{\!(0)}=\vert\Psiv\rangle\langle\Psiv\vert \, \label{32}, \fe
and whichever {\it a posteriori} probability operator 
\be {\Pbr}_{[\ivr]}=\Pr_\ivr^{-1}\vert\Psiv_\ivr\rangle\langle\
Psiv_\ivr\vert\, ,\label{33}\fe
 may turn out to apply.

If the presumption that the system was initially in a pure state is 
replaced by the more general supposition that it was in an initial state 
describable by an {\it a priori} probability operator, ${\Pbr}_{\!(0)}$ 
say, consisting of an arbitrary sum of pure state operators, then by 
considering the effect on each member of such a sum it can be seen that 
the effect of the first step of the observation process described in the 
preceding paragraph will be to provide a provisional probability operator 
given no longer by the simple formula (\ref{31}) but by the more general 
prescription
\be {\Pbr}= \sum_\ivr \Pr_\ivr\, {\Pbr}_{[\ivr]}\, ,\label{35}\fe
that is known~\cite{Bub} as Luder's rule, in which the operators
${\Pbr}_{[\ivr]}$ are the {\it a postiori} probabilities for the relevant 
output channels, i.e. the relevant eventualities $\evr_\ivr$, which are what 
Everett referred to as ``branches''. In accordance with the formula 
(\ref{24}), these {\it a posteriori} probability operators, and the 
corresponding probabilities, are given in terms of the {\it a priori} 
probability operator ${\Pbr}_{(0)}$ by
\be {\Pbr}_{[\ivr]}=\Pr_\ivr^{\,-1}{\ebr}_\ivr {\Pbr}_{\!(0)}{\ebr}_\ivr\, ,
\hskip 1 cm \Pr_\ivr={\rm tr}\{{\Pbr}_{\!(0)}{\ebr}_\ivr\}\, ,\label{36}\fe
and it is to be noted that they are also recoverable, using expressions 
of the same form 
\be {\Pbr}_{[\ivr]}=\Pr_\ivr^{\,-1}{\ebr}_\ivr {\Pbr}\,{\ebr}_\ivr\, ,
\hskip 1 cm \Pr_\ivr={\rm tr}\{{\Pbr}\,{\ebr}_\ivr\}\, ,\label{37}\fe
from the ensuing provisional probability operator (\ref{35}).

\bigskip
{\bf 8.  The deficiency of the Everett  interpretation}
\medskip

Before exposing the essential deficiency of the Everett doctrine, I would 
like to rectify an accessory misconception to which it has given rise. 
In its usual presentation, the use of the term branch is motivated by the 
notion that the number of relevant channels increases whenever an 
observation is made. It is important to recognise that this idea -- of 
perpetual multiplication of the relevant number of branch-channels --
is, as a general rule, misguided.  It is based on the -- rarely realistic
-- presumption that the {\it a priori} state of the system under
consideration is pure, consisting just of a single branch-channel, whereas 
in a generic case (for the reasons discussed above) ${\Pbr}_{\!(0)}$
will already be mixed, involving as many branch channels as ${\Pbr}$,
so no actual increase occurs. It is thus more appropriate as a metaphore 
to speak of channels rather than branches, which is why I have chosen, 
as a compromise, to use the term branch-channel. In any case (even if the 
initial state really was pure) the commonly accepted idea that -- as more 
and more information is obtained by successive observations --  the number 
of branches will go on increasing is also unrealistic for a different 
reason, which is that a given finite system cannot continue to acquire more 
and more information without limit. After a certain amount of information 
has been acquired, the system will saturate, so that further information 
will be able to be taken into account only by a (Landauer 
type~\cite{Vedral}) process involving the erasure of previously recorded 
information in order to release the necessary memory space. The number of 
channels available for useful observation can at best be only a small 
fraction of the number of dimensions needed for a complete physical 
representation of the sensor, which in practice (if he, she, or it is a 
system constituted from a finite number of molecules with a finite total
energy in a finite volume) will of course be limited.

Bounded though it must be, the number of branch-channels -- meaning the 
number of eventualities that may be observationally distinguished -- in a 
given (human or other) sensor system can indeed be very large. It is 
this consideration that has exposed the Everett proposal~\cite{Everett57}  
that all the branches are ``actual, none any more real than the rest'' to 
the criticism~\cite{Leslie83} that it entails a ``bloated ontology''. 
However, as I have previously remarked~\cite{Carter93} as far as the 
scientific desideratum of Ockham's razor (meaning economy of formulation) 
is concerned it does not matter how extensive or otherwise the ensuing 
``ontology'' may be.

A more serious reason for dissatisfaction with the Everett doctrine of 
quantum theory is its failure to apply its own declared rules in a 
coherent manner, which has made the question of the interpretation of
this ``interpretation'' the subject of much discussion~\cite{DeWitt, Graham}.
The assertion that the branches are ``actual'' seems to imply their 
ontological reality, but Everett's categorical denial that any one is 
``more real than the rest'' is followed by the Orwellian 
admission~\cite{Everett57} in a subsequent  paragraph that ``in order to 
obtain quantitative results'' the branches must be given ``some sort of 
quantitative measure (weighting)''.

The aim of the Everett program, as expressed by De Witt~\cite{DeWitt}, is 
to construct a theory ``in which it makes sense to talk about the state 
wector of the whole universe. This vector never collapses, and hence the 
universe as a whole is deterministic''. The troublesome 
problem~\cite{Deutsch,Wallace,Greaves} is how to use such an ultimately 
deterministic model to obtain the probabilistic predictions that work so 
well in local applications of quantum theory. As Graham~\cite{Graham} puts 
it ``Everett attempts to escape from this dilemma by introducing a 
numerical weight for each world''.The work of Graham and of 
Hartle~\cite{Hartle} has shown that Everett's ``weighting'' scheme does 
successfully reproduce the usual probabilistic predictions, so much so that 
indeed the distinction between the terms ``weighting'' and ``probability'' 
can be seen to be merely semantic. Changing its name  to ``weighting'' (or 
``propensity'', which is another traditionally favoured alternative) does 
not solve the problem of interpreting the meaning of the ``probability'' 
that is involved. 

It is clear that Everett and his followers have so far failed to
achieve their declared objective. Their bold attempt to solve the -- 
originally local -- interpretation problem by reintroducing determinism
at a global level has been helpful for providing a deeper understanding 
of many of the issues involved, but the question of how much ``reality'' 
should be attributed to the probabilistically ``weighted'' branch-channels
has nevertheless remained unsolved until now.

My purpose here is to present a recent clarification~\cite{Carter03} whereby 
this issue is not so much decided as transcended, in conformity with the 
precept that questions of ontology are of a theological nature that is 
beyond the scope of ordinary science (whose modest ambition is to account 
for appearances, and not for ultimate reality, whatever that may mean). The 
anthropic approach described below provides a framework in which an 
intellectually coherent interpretation can be provided in a manner that 
leaves plenty of scope for adjustment, and that is compatible not only with 
an (unbloated) ``oriental'' option, in which hardly any of the relevant 
branches need be considered to be ``real'', but also with a (scientifically 
indistinguishable, but theologically very different) ```occidental'' option 
in which they might all be describable as ``actual''.

\bigskip
{\bf 9. The side issue of the provisional distribution.}
\medskip

Whereas zealous adherents of the Everett doctrine --  and {\it a fortiori}
of the naive version of the Copenhagen interpretation that was discussed 
above -- would have it that some sort of objective reality can be 
attributed to the state vector on a sufficiently large scale, and hence 
to the probability operator that would be relevant on a more local scale, 
on the other hand most other schools of thought, including less naive 
versions of the dualistic Copenhagen interpretation, would concur with 
the supposition adopted here to the effect that such entities are 
essentially of a subjective nature. This contrasts with the status of 
the Hilbert space operator algebra of  eventualities and observables, 
which have a more objectively well defined nature. According to this
principle, the amplitudes (and corresponding ``weightings'') of Everett 
type ``branches'' should be considered as ultimately subjective, whereas 
the branches themselves can be considered to be objective -- which does 
not of course entail that such mathematical structures are ``real'' in 
any ontological sense. 

Before leaving the subject of the ``branching'' process (misnamed because 
the number of branches involved in the description of a subsystem need not 
increase, and might even decrease, when an interaction occurs) it is worth 
commenting further on the nature of the process whereby an {\it a priori} 
probability operator ${\Pbr}_{\!(0)}$ is replaced by the corresponding 
provisional probability operator ${\Pbr}$ as given by (\ref{35}) and 
(\ref{36}). The original discussions of this process were formulated in 
terms of what Dirac~\cite{Dirac} referred to as the Schroedinger picture, 
wherein states are considered to have a time dependence whereby the 
evolution from an initial time $t_{(0)}$ say to a later time  $t$ is 
given by an operator transformation ${\Pbr}_{\!(0)}\ \mapsto\ {\Pbr}$ 
that will be given, in the special case of a pure state for an isolated 
system, by a corresponding vector transformation  $\vert\Psiv_{(0)} 
\rangle\ \mapsto\ \vert\Psiv\rangle$. In the special case of an 
isolated system, such a transformation will be given by a unitary operator 
${\bf U}$ (that is continuously generated by some Hermitean Hamiltonian) 
according to prescriptions of the standard form $\vert\Psiv\rangle= {\bf U}
\vert\Psiv_{(0)}\rangle$ and ${\Pbr}={\bf U}{\Pbr}_{\!(0)}{\bf U}^{-1}$. 
However the transformation will in general be of a less simple 
(non-unitary) type when interaction with an external system is involved. 
The idea, as discussed by von Neumann, was that the preparation of an 
actual experimental observation should involve an arrangement whereby a 
transformation of this latter (non-unitary) type produced a provisional 
probability operator ${\Pbr}$ of the required form, as given by the 
Luder formula (\ref{35}).

As originally pointed out by Dirac~\cite{Dirac}, a representation in terms 
of such a Schroedinger picture can be translated into an equivalent
representation in terms of the kind of Heisenberg picture that has been 
implicitly adopted throughout the present discussion. In this kind of 
representation, the relevant state vector $\vert\Psiv\rangle$ or probability 
operator ${\Pbr}$ is considered to be time independent, and the effect 
of Schroedinger type time translations is allowed for by corresponding 
transformations of the relevant observables and their constituent 
eventualities. In the special case of an isolated system these 
transformations will be of the standard unitary type, so that for example 
if ${\ebr}_{(0)}$ is the projection operator corresponding to some 
particular eventuality at a time $t_{(0)}$ then the corresponding time 
transposed eventuality at a later time $t$ will be given by 
\be {\ebr}={\bf U}^{-1}{\ebr}_{(0)}{\bf U}\, .\label{41}\fe
The essential advantage of using a picture of this kind is that there 
is no impediment to its extension to (General Relativistic and other)
applications for which no globally well define Newtonian type time 
parametrisation may be available, so that the concept of a time translation 
relation of the form ${\ebr}_{(0)}\mapsto {\ebr}$ might make sense only
for very particular locally related eventualities.

As seen from this Heisenberg (as opposed to Schroedinger) point of view,
the process of preparation of an experimental observation in the manner 
prescribed by von Neumann should be thought of, not as the replacement of 
an {\it a priori} probability operator ${\Pbr}_{\!(0)}$ by a different 
provisional probability operator ${\Pbr}$, but as the the replacement of 
an initially envisaged, but perhaps maladapted, observable 
$\{{\ebr}_{(0)}\}$ by an appropriately adjusted observable $\{{\ebr}\}$ 
with respect to which the probability distribution already has the required 
Luderian form (\ref{35}).  

From this point of view, there is no need to bother about any distinction 
between {\it a priori} and provisional probability operators (which -- 
in view of the possibility of using  (\ref{37}) instead of (\ref{36})
-- were in any case equivalent for the practical observational purpose
under consideration). What matters for the purpose of making what von 
Neumann would consider to be a satisfactory observation is the choice of 
a suitably adjusted  observable ${\ebr}$. However the main point I wish 
to emphasize at this stage is that although it may be of technical interest 
in particular applications, the importance of the issue of obtaining a 
satisfactory observation in the sense specified by Luder's rule has been 
greatly exaggerated, in so far as its relevance to the ultimate 
interpretation of the meaning of the observations process is concerned.  
To start with there is the consideration that the Luderian desideratum is
obtainable not only by the non-trivial process described above, whereby 
$\{{\ebr}\}$ is adjusted to a previously chosen probability operator, 
but also by the trivial process whereby the subjective {\it a priori} 
choice of ${\Pbr}$ is adjusted {\it ad hoc} to fit a prescribed observable 
$\{{\ebr}\}$, an adjustment that in no way diminishes the credibility of 
its implications, as can be seen from the equivalence of the prescriptions 
(\ref{36}) and (\ref{37}).
  
A more fundamental reason why the question of the Luderian transition
is irrelevant is that when an observation has been actually carried out 
(not merely planned) one will be left just with a single confirmed 
eventuality, $\evr_\ivr$. Such a single eventuality might be incorporated 
with others to constitute a complete observable set (spanning the entire 
Hilbert space) in many different ways, whose substitution in the Luder 
formula (\ref{35}) would provide many different results. Nevertheless, 
however that might be, and regardless of any distinction that may or may 
not have been made between an {\it a priori} probability distribution 
${\Pbr}_{(0)}$ and a provisional probability ${\Pbr}$, one will be left 
with an unambiguously specified {\it a posteriori} probability distribution
${\Pbr}_{[\ivr]}$, which is all that matters for the purpose of
subsequent predictions one may wish to make. 

The upshot is quite simply that someone (such as Wigner when concerned 
about Shroedinger's cat) should use the {\it a posterior} distribution 
when the relevant information has become available, and until then 
should just continue to use the ordinary {\it a priori} distribution. 
One should avoid getting sidetracked (as so many of Everett's followers
 have been) by intermediate Luderian technicalities, whose analysis is of 
little relevance to the two outstanding issues that remain. In addition 
to the question of interpretation, which will be addressed from an 
anthropic point of view below, the other outstanding issue is of course 
the usual practical Bayesian dilemma of how to decide quantitatively 
what {\it a priori} distribution should be used in a particular context -- 
something that can sometimes be resolved just by symmetry considerations 
(as in the coin tossing example described above).

\bigskip
{\bf 10. Perceptions and perceptibles.}
\medskip

An important idea that was latent in much of the preceding discussion 
is that some privileged eventualities and observables are more naturally 
significant than others. 

In the discussion of Luder's rule it was remarked that this rule can be 
interpreted as selecting a privileged class of observables, but I would 
emphasize before continuing that privilege of that kind is not what I am 
concerned with here, because it is ultimately dependent on an arbitrary 
subjective choice of the relevant {\it a priori} probability distribution. 

The kind of privilege I am concerned with here is something that depends on 
the essential nature of the system under consideration in a manner that is 
independent of the choice of the probability distribution. This is something 
that could be said about Bohm's idea~\cite{Bub} of privileging position 
with respect to its dynamical conjugate, namely momentum, but that 
particular choice is something that would not seem very natural to the 
numerous physicists whose mental life is based in Fourier space.
 
The kind of privilege that seems to me more relevant for the interpretation 
question is something that would be rather generally recognised as being 
imposed by the circumstances in particular cases. It is exemplified most 
simply by the existence of a privileged choice (determined by the 
background magnetic field) for the the particular spin eventualities 
characterised as ``up'' and ``down'' in the Stern Gerlach experiment that 
has been discussed above. It is exemplified by many familiar kinds of 
apparatus, such as can be found in scientific laboratories, and increasingly 
in ordinary homes, whose output is typically presented in terms of what, at 
the highest resolution usually turns out to consist of simple integer valued 
observables, such as the alternative eventualities in the range from 0 to 9 
for a digit in a counter output, or the binary alternatives for a particular 
pixel on a screen to be ``on'' or ``off''. It is mathematically possible to 
use other bases for a Hilbert space description of such systems, for example 
by working with eventualities defined as linear superpositions of ``on'' and 
``off'' states of screen pixels, but that is evidently not the kind of 
treatment for which such an apparatus was intended by its designer.  

Although the degree of complexity of the systems involved is very different, 
it seems to me that there is a rather strong analogy between the special 
role of the ``on'' and ``off'' states for a pixel on a screen and the 
``awake'' and ``sleeping'' states of Schroedinger's cat. The privileged 
status of the particular eventualities in question can be accounted for as 
the result of a process of design that is attributable in the first case, 
not just to an individual engineer, but to the collective activity of a 
scientific community, while it is attributable in the second case to a very 
long history of biological evolution by Darwinian selection. Having said 
this about the cat, the next thing to be said is of course that the same 
applies to Schroedinger and Wigner, for whom the relevant privileged 
eventualities are states of mind corresponding to the realisation that the 
cat is awake or not as the case may be. 

Whatever doubts we may have about the status of the cat, we must recognise 
that Schroedinger and Wigner are closely analogous to ourselves (meaning 
the author and presumed readers of this essay) which means that insight 
into the working of their minds can be obtained from our own experience. 
Since the only eventualities about whose reality we can be sure are the 
conscious perceptions in our own minds (of which some, namely those 
occurring in dreams, are evidently uncorrelated with anything outside) 
corresponding to the ``mind states'' whose essential role has been 
recognised by several authors, such as  Donald~\cite{Donald}, 
Lockwood~\cite{Lockwood}, and in particular by Page~\cite{Page}, whose line 
of approach is followed here. It seems reasonable to postulate the validity 
of Page's principle according to which conscious perceptions are the only 
eventualities that can be considered to actually happen. It also seems 
reasonable to make the concomitant postulate that these perceptions must 
belong to some restricted class of privileged eventualities of the kind 
discussed in the preceding paragraph. I shall refer to the eventualities of 
this subclass as perceptibles.

In his ``sensible quantum theory''~\cite{Page}, Page has attributed a 
privileged role to a class of observables that  he refers to as ``awareness 
operators'', which I interpret to  mean observables whose individual 
constituent eventualities are the perceptibles introduced in the previous 
paragraph. Page has used these particular operators to develop a refined 
version of the Everett interpretation, in which the branches -- or as I 
would prefer to say, channels -- that matter are specified with respect to 
these awareness operators.  Thus whereas Everett's original version might 
attribute ``actuality'' to branches defined with respect to observables of 
a rather arbitrary kind, Page's more refined version would attribute 
``actuality'' only to branches of an appropriately restricted kind, namely 
the channels that are specified by perceptibles. Having thus provided a 
much clearer idea of which channels are actually needed, Page was still 
left with the problem of interpreting what, following Everett's evasive 
example, he referred to as their ``weighting''. The point at which Everett 
stumbled was in trying to reconcile his recognition that the weighting was 
needed with his preceding claim that all the branches were equally real. 
Page came up against the same problem with respect to the claim to the 
effect that all the perceptibles are actually perceived.

\bigskip
{\bf 11. The anthropic abstraction}
\medskip

A corresponding paradox is reached from a rather different angle in the 
approach I am developing here, which is in agreement with that of 
Page~\cite{Page} in so far as the special role of perceptions is concerned, 
but differs in affirming that the weighting in question must be considered 
to have an essentially subjective and probabilistic nature. The intrinsically 
probabilistic nature of  models of the kind advocated here raises the 
problem of what it can mean to attach a probability to the actuality of an 
eventuality in the mind of someone else if the only events one can 
actually observe are are those occurring in one's own mind.

Before presenting what I think is the only acceptable way of dealing with 
this paradoxical problem, I would mention two less satisfactory ways of 
resolving the issue that have been suggested in the past. The first way is 
of course that of the solipsist, who would deny the existence of any 
conscious perceptions other than his (or her~\cite{Greaves}) own, with the 
implication that the apparent analogy between oneself and others such as 
Schroedinger is merely a superficial illusion. The second way (which unlike 
that of the solipsist has been followed up deliberately by many physicists, 
starting with de Broglie) is to revert to a deterministic description of 
the world, providing a theoretically well defined answer to the question of 
what really happens by denying the (experimentally well established) 
validity of the essentially probabilistic description provided by orthodox 
quantum theory. Neither the first nor the second of these ways of solving 
the problem can be said to actually resolve the paradox: they merely evade 
the issue by dropping one or other of the essential (experimentally 
motivated) elements of the problem, which is that of providing an 
essentially probabilistic treatment of perceived reality that respects 
the apparent symmetry between different people.

A historical analogy is provided by the incompatibility between Maxwellian 
electromagnetism and Newtonian gravity, which was ultimately resolved by 
their unification in Einstein's General Relativity. The problem to be dealt 
with here is that of reconciling subjective probability with objective 
reality. The only way that I know of for solving this problem in a 
satisfactory manner is the anthropic approach, which faces the issue head 
on~\cite{Carter03} without denying the validity of the considerations that 
lead to the paradox.

It is worth emphasising, by the way, that the problem is not specifically a 
problem of quantum theory, but also arises in probabilistic versions of 
classical theory, as was recognised, I suspect, by many of those who were 
hostile to anything associated with the name of Bayes. The importance in 
this context of the quantum revolution is that it changed the status of 
Bayesian theorists from that of radicals (because they were willing to 
abandon determinism) to that of reactionaries (because they continued to 
use old fashioned Boolean logic).

The situation, as I understand it,  is as follows. Suppose that to describe 
a system that includes ourselves (but, for the sake of finiteness, perhaps 
not the whole of the universe) we have set up  some (classical or quantum) 
theory that provides probabilities for an extensive class of eventualities. 
This class includes a specially privileged subclass of eventualities that 
I shall refer to as perceptibles, which are the only ones that can be 
actually observed as conscious perceptions. The set of such perceptions (not 
just yours and mine, but also those of everyone else) can be described as 
objective, and it is the only thing in the theory that can be considered to 
be real

You have an objective model attributing probabilities to perceptibles, not 
only your own but those of other people. But what sense can it make to 
attribute a probability to an observation you cannot make? If you are 
Wigner, what sense can it make -- even in a classical theory -- to use an 
objective distribution attributing probability to something that can can 
only be known by Schroedinger? The contradiction arises when Schroedinger 
makes the Bayesian transition to the relevant {\it a posteriori} 
distribution, while Wigner continues for the time being to use the {\it a 
priori} distribution. How in these conditions can either of these 
distributions be considered to be objective?

The resolution to this paradox is provided by what may be called the 
anthropic abstraction (so called because it underlies of what I designated 
-- perhaps inaptly -- as the anthropic principle~\cite{Bostrom}).  The 
paradox that arises in this case (as in many others) can be attributed to 
an unnecessary assumption that has been consciously or subconsciously taken 
for granted. The unnecessary assumption is that of knowing in advance who 
one is. The {\it anthropic abstraction} consists in refraining from 
assuming in advance that one has the identity of some particular sensorial 
observer in the model, so that one's status {\it a priori} is that what I 
shall refer to as an abstract {\it perceptor}.  It is not until the actual 
happening of the perception that one can know whether one is Schroedinger, 
or Wigner, or whoever else may be included in the model. 

It is of course to be understood that the perceptible eventualities that 
are involved in this anthropic approach cannot just be of the elementary 
type exemplified by the observation that someone else is awake, but need to 
include eventualities of the more complicated kind known as {\it consistent 
histories}~\cite{Gell}. The sort of eventuality that needs to be envisaged 
is not simply that of finding oneself to be Schroedinger, but that of 
finding oneself to be Schroedinger at a particular instant in his life, 
with all the memories he would have had at that moment.

The use (which I see no satisfactory way of avoiding without reverting to 
determinism) of the anthropic abstraction entails the need to adopt some 
kind of anthropic principle, by which I mean some kind of prescription for 
attributing appropriate probabilities to the relevant perceptible 
eventualities. The rather crude kind of anthropic principle that I have put 
forward on previous occasions~\cite{Carter83} was concerned with the 
attribution of probability to entire observer systems, (such as those 
associated with the names of Schroedinger or Wigner) without getting into 
the details of particular moments in their lives. For the applications I was 
then considering, it was sufficient to use a crude statistical treatment 
attributing equal weight to all terrestrial or extraterrestrial observers 
who can be considered to be sufficiently like ourselves to be describable as 
``anthropic''. However -- as several authors have already 
remarked~\cite{Page, Bostrom, Leslie96} -- the more detailed applications I 
have been considering here (particularly those involving quantum effects) 
require the use of a more refined kind of anthropic 
principle~\cite{Carter03} that will distinguish not just between anthropic 
individuals but between different instants in the lives of such individuals. 

The question that naturally arises at this point in this line of reasoning 
is whether it can suffice to use just the probability weightings that  are 
directly provided by orthodox quantum theory (such as has been discussed 
above) in conjunction with some prescription for deciding which of the 
many mathematically defined eventualities in the model should be considered 
to have the privileged status of perceptibility?

\bigskip
{\bf 12. Uniqueness of the perceptor?}
\medskip

In the subsequent subsections I shall  address the scientifically important 
question of the attribution of the required anthropic probability. However 
before doing so, I would like to digress by mentioning another question of 
a less scientific nature that might also be a subject of philosophical 
discussion in the future.

This is the question of the nature of what I have referred to as a 
perceptor, whose actual perceptions are the only entities within the model 
that are considered to be real (which is not to deny the reality, in some 
theological sense, of other entities beyond the scope of the model).  The 
perceptor acquires an a posteriori identity (of an ephemeral nature) as a 
material observer (such as Schroedinger) on the occasion of an actual 
perception, but what about the immaterial identity the perceptor might 
have {\it a priori}?

Is the perceptor unique? The notion that all anthropic observers might just 
be avatars of a single perceptor will not seem strange to anyone familiar 
with oriental (Hindu-Buddhist) religious tradition. (A scientific analogy 
that comes to mind is Feynmann's idea that the universe is inhabited only 
by a single electron, which is able to follow all the world lines that we 
usually attribute to distinct elections by also following -- but in a time 
reversed sense -- the other world lines that we attribute to positrons.) 
The obvious Wheelerian epithet for the succinct encapsulation of this idea  
-- namely that we all share the same abstract identity --is {\it solipsism 
without solipsism}.

The postulate of a unique perceptor has the advantage of being particularly 
economical in the sense required by Ockham's razor. Nevertheless, in the 
framework of the occidental (Judaeo-Christian-Islamic) religious tradition 
it might seem more natural to suppose that there are many distinct 
perceptors. What is not permissible, however tempting it may seem, is to 
suppose that distinct perceptors are correlated with distinct anthropic 
observers, such as Shroedinger and Wigner: the essence of the anthropic 
abstracion is that a perceptor has the potential for actualisation in any 
observer state that has a non zero probability amplitude.  The only way you, 
as a material observer, can claim an exclusive monopoly of the potential 
for actualisation of your own perceptor, is by adopting an {\it a priori} 
probability distribution that attributes no weight to anyone other than 
yourself, in other words by adopting the (unnaceptable) 
autocentric attitude describable as solipsism {\it with} solipsism.

For someone whose objection to the Everett doctine was based not on not on 
its failure to follow its own declared rules, but on the ontological 
bloating~\cite{Leslie83} implicit in the many universe doctrine, the 
present idea that one might adopt a many perceptor doctrine might be felt 
to even worse. Whereas the number of Everett branch-channels is restricted, 
as I have remarked above, by the limited information content for any 
finite system, on the other hand there is no limitation at all on the 
number of distinct perceptors that might be conceived to exist, and that 
might all have a chance of undergoing the experience of being Schroedinger 
at some moment in his life. 

The idea that there might be an unlimited number of distinct perceptors 
may be abhorrent to anyone for whom ontological economy is a desideratum, 
but on the other hand it might be extremely attractive to those who still 
hanker after determinism. Indeed for those who consider that in order to be 
meaningful the concept of probability must be defined in terms of 
frequencies of the outcome of many identical performances of the same 
experiment, the many perceptor doctrine can provide what is desired. If the 
number of perceptors is vastly larger than the number of anthropic 
observers in the model, then each observer state (even those that are 
relatively improbable) would actually be perceived by a large number 
(albeit a small fraction) of the perceptors. This would provide the 
desired frequency interpretation for the probability distribution. By using 
the anthropic abstraction in this ontologically uninhibited manner, it is 
at last possible to deliver what the Everett program sought, which may be 
epitomised as {\it probability without probability}.

Multiplication of the number of sensors is not the only way of obtaining 
probability without probability, if that is what is desired. Another number 
whose magnification can achieve the same result is the number of 
perceptions that each particular perceptor is allowed to make. The 
supposition that there is a large number of perceptors, each allowed to 
make only a small number of perceptions or, even restricted to a single 
perception, is ontologically equivalent to the supposition that there is 
just a single perceptor who is allowed to make a large number of 
perceptions. As far as ontology is concerned, all that counts is the total 
number of perceptions.

Whether -- as in the oriental version of the anthropic interpretation -- 
there is a unique perceptor, or whether -- as in the occidental version -- 
the number of perceptors is large (even compared with the number of 
anthropic observers) -- is an issue that belongs to the realm of theology 
rather than science. The same can be said about the (more ontologically 
relevant) number of total perceptions, which may seem important to those 
who believe in probability only when formulated in terms of frequencies, 
but which in no way affects the way the theory is actually applied in 
practice. All that matters for scientific purposes is the relative 
probablity distribution for the perceptions, which will now be discussed.

\bigskip
{\bf 13. Anthropic weighting: the proper ansatz?}
\medskip
 
On the basis of what precedes, it seems reasonable to suppose that, from  
the point of view of a perceptor, the ``net'' probability, ${\PP}$ say, of 
a particular perception $\evr_\ivr$ within a particular subsystem 
(representing the part of the universe under consideration) should be 
given by an expression of the form 
\be \PP\{\evr_\ivr\}=\calP_\evr\, \Pr_{[\calO]}\{\evr_\ivr\}
\, ,\label{50}\fe
where $\Pr_{[\calO]}\{\evr_\ivr\}$ is the ordinary `ègross'' classical or 
quantum mechanical probability (as calculated in the manner described above) 
for the particular perceptible eventuality $\evr_\ivr$ to occur on the 
occasion when the relevant Page type awareness observable $\{\evr\}$, is 
actually observed, while the ${\calP}_\evr$ is the anthropic factor giving 
the probability for the perception to belong to that particular observable 
set. A sensor of the familiar macroscopic but localised kind -- exemplified 
by an ordinary computer, or a human observer -- will be characterisable by 
a fairly well defined world line with a proper time parametrisation $\tau$, 
in terms of which the anthropic probability factor will be expressible in 
the form
\be {\calP}_\evr=\dot{\calP}\,\Delta_\evr\tau\, ,\label{52}\fe
where $\Delta_\evr\tau$ is the relevant proper time duration, and 
$\dot{\calP}$ is a corresponding probability rate factor, whose integral 
\be {\calP}=\int \dot{\calP}\,{\rm d}\tau \, ,\label{54}\fe
will be interpretable as the giving the total probability for the 
perception to occur in at some stage in the life of that particular observer. 

Whereas the conditional probability designated by a roman capital $\Pr$ in 
(\ref{50}) is of the ordinary  kind that is provided by the relevant 
classical or quantum physical theory for  the system under consideration, 
on the other hand the anthropic probability factor designated by a 
caligraphic ${\calP}$ (which is also conditional in so much as it is 
subject to the condition of restriction to that particular system within 
the universe) can only be provided by what I call an {\it anthropic principle}.

In my earlier discussions of applications~\cite{Carter83} that were not 
concerned with discrimination between individuals, but with averages over 
entire populations, it was good enough to suppose that provided they were 
sufficiently similar to ourselves (that was the motivation the -- rather 
debatable -- choice of the term anthropic) the relevant total probability 
${\calP}$ per observer could be taken to be the {\it same} for each one, in 
accordance with what Vilenkin has referred to as a postulate of {\it 
mediocrity}, and what I would refer to just as a postulate of {\it approximate 
symmetry}. The application of such a mediocrity postulate in the present 
context gave rise to what I called the {\it weak anthropic principle}, whose 
purport is that the anthropic probability factor should take a fixed value 
\be {\calP}={1\over N}\, ,\label{56}\fe
where $N$ is the number of anthropic observers that come into existence 
within the system under consideration (so that if the system were scaled up 
to include a larger chunk of the universe, with a larger population number 
$N$, then the value of ${\calP}$ would be correspondingly scaled down.)

The ordinary (weak) anthropic principle formulated in the preceding 
paragraph will evidently not be enough for more detailed purposes, such as 
comparison of the probability of finding oneself to be someone very short 
lived (as in the case of a child that dies in infancy) with that of finding 
oneself to be someone more long lived (as in the case of a normal adult). 
For such a purpose, the most naively obvious possibility is to adopt the 
ansatz what I would call the {\it proper} anthropic principle, meaning the 
postulate of a fixed universal value for the anthropic probability rate 
${\calP}$, which would be given numerically by
 \be \dot{\calP}={1\over \langle\tau\rangle N}\, ,\label{58}\fe
where $\langle \tau\rangle$ is the average total proper lifetime of an 
anthropic observer in the system. In so far as the total probability over 
the total lifetime $\tau$ of an observer is concerned, adoption of the 
proper anthropic principle (\ref{58}) evidently entails that (\ref{56}) 
should be replaced by
\be {\calP}={\tau\over\langle\tau\rangle N}\, .\label{60}\fe

The foregoing proper refinement of the original anthropic principle 
(\ref{56}) should, I think, be good enough for a wide range of applications. 
However for the purpose of comparing observers of very different kinds (for 
which the qualification anthropic might not be so appropriate) such as 
extraterrestrials and cats, not to mention babies in our own species, the 
plausibility of (\ref{58}) is much less obvious.

\bigskip
{\bf 14. Micro-anthropic principle: the entropic ansatz.}
\medskip

A hint toward a more plausible (though not so easily applicable) 
alternative is discernible in the  response to the eschatological problem 
posed by Islam~\cite{Islam77} that was provided by Dyson, who 
suggested~\cite{Dyson79} that what really matters is not the proper time 
duration of an interval but how much information is effectively processed 
therein. There is of course room for discussion about how to quantify what 
is effectively processed (as opposed to what is merely stored in a memory) 
in the case even of an ordinary computer and hence much more so for in the 
case of a feline or human mind. Estimating that the duration of a human 
``moment of consciousness'', which presumably corresponds to what is 
denoted here by $\Delta_\evr\tau$ has the same order of magnitude as was 
supposed in the more recent work of Page~\cite{Page}, namely a significant 
fraction of a second, Dyson deduced (from the fact that a the heat 
production of an entire  human body is typically about 200 Watts at a 
temperature of 300$^o$ K) that the corresponding entropy production, 
$Q_\evr$ say,  is of the order of $10^{23}$ bits. However experience with 
the analogous problem for computers indicates that the amount of 
information $S_\evr$ that can be judged to have been effectively processed 
by the mind itself during the corresponding period of perception -- and the 
associated Landauer entropy production~\cite{Vedral} -- must have a vastly 
smaller value $S_\evr\ll Q_\evr$ that is not so easy to evaluate.

A plausible prescription for the evaluation of the processed information 
$S_\evr$ will however be available if we have a sufficiently detailed 
(quantum not just classical) theory to characterise the Hilbert space 
projection operator ${\ebr}_\ivr$ corresponding to a particular perception 
$\evr_\ivr$ under consideration. If we suppose that this particular 
perception belongs to a complete set of eventualities having the same rank 
(i.e. subspace dimension) $\calR_\evr={\rm tr}\{{\ebr}_\ivr\}$ constituting 
an observable $\{\evr\}$ in a Hilbert  space of dimension $\calN={\rm tr}
\{ {\Ibr}\}$, so that the corresponding number of Everett type branch 
channels is $\calN_{\!\evr}=\calN/\calR_\evr$, then the associated 
information capacity will be given by 
\be S_\evr={\rm log}\{\calN_{\!\evr}\}={\rm log}\{{\rm tr}\{ {\Ibr}\}\}
-{\rm log}\{ {\rm tr}\{ {\ebr}_\ivr\}\}\, ,\label{70}\fe 
using a logarithm with base 2 if one wants to use Shannon's bit units, or 
using a natural (Naperian) logarithm if one wants to use the entropy units 
that are commonly preferred by physicists. This information capacity 
represents the maximum amount of information that can be given -- for a 
probability distribution $\Pr_\ivr$ ($\ivr=\oner, ..., \calN_{\!\evr}$) -- 
by Shannon's formula $S=-\sum_\ivr \Pr_\ivr\,{\rm log}\{\Pr_\ivr\}$.

What I would propose is that the formula (\ref{70}) be used as an estimate 
of the amount of information that can be considered to be processed during 
the perception $\evr_\ivr$, and that the corresponding anthropic 
probability should be postulated to be proportional to this, i.e. the 
required factor in (\ref{50}) should be taken to be given by
\be {\calP}_\evr=\alpha S_\evr\, ,\label{72}\fe
where $\alpha$ is a fixed proportionality factor that is chosen so as to 
ensure satisfaction of the usual requirement that the total probability  
(over all the relevant world lines) should add up to unity. According to 
this {\it micro-anthropic} principle -- which might appropriately be 
described by the term {\it entropic} principle --  the probability rate 
factor will not have a fixed value (as was postulated by the proper 
anthropic principle formulated above) but will be given by
\be \dot{\calP}=\alpha {S_\evr\over\Delta_\evr\tau}\, .\label{73}\fe

The advantage of using the term entropic principle for this ansatz is that 
it emphasizes its virtue of being applicable in principle not just to 
observers qualifiable as anthropic, in the sense of being sufficiently 
similar to ordinary adult humans,  but also to very different kinds ranging 
from such familiar examples as babies and cats to the highly exotic 
extraterrestrial observers whose survival at extremely low temperatures was 
envisaged by Dyson. A rather obvious application of this entropic principle 
is its use as evidence against Dyson's conjecture~\cite{Dyson79} that 
civilisations constituted by observers capable of surviving at the 
extremely low temperatures predicted~\cite{Islam77} for a non compact 
universe in the far distant future would be able to survive indefinitely 
with respect not just to proper time but with respect to the relevant 
information processing measure. If this conjecture were correct it would 
mean that the probability measure defined according to (\ref{73}) by the 
entropic principle would diverge toward the future. The contrary 
prediction by Islam~\cite{Islam77} that ``it is unlikely that civilisation 
in any form can survive indefinitely'' is therefore overwhelmingly favoured 
by the fact that we do not observe ourselves to be incarnated in an 
asymptotically viable low temperature life forms (if any such can exist at 
all) but in carbon based life forms adapted to a (cosmologically ephemeral) 
conditions of moderate temperature. 

It is to be emphasised that the preceding argument against the likelihood 
of long term survival is entirely dependent on the acceptance of the kind 
of {\it a priori} probability distribution proposed (as a matter of choice, 
not merely as a tautology) by the anthropic principle and its entropic 
extension.  Dyson's writing in this and other analogous contexts -- notably 
that of the prospects for our own terrestrial civilisation in 
particular~\cite{Leslie96} -- give the impression that he personally 
prefers an {\it a priori} probability distribution of the traditional kind 
based on what I would refer to as an autocentric (or preordination) 
principle, to the effect that the attribution of non zero weighting should 
be restricted retroactively to wherever one already finds oneself to be. 
Although it may be logically admissible as an alternative to principles of 
the anthropic kind, I would maintain that such an autocentric attitude is 
scientifically unreasonable, in so much as it violates the desideratum that 
comparable observers be treated objectively on the same footing. By 
adopting such an attitude~\cite{Dyson96}, Dyson implicitly assumes for 
himself a privileged position to which other observers (such as Wigner and 
Schroedinger) are not admitted.

Before leaving the subject of logically admissible (even if not 
scientifically reasonable) alternatives to principles of the anthropic 
kind, I would mention a conceptually possible alternative that is quite the 
opposite of the autocentric deviation described in the previous paragraph. 
Instead of prescribing an {\it a priori} probability distribution with 
weighting restricted to material observers as in the anthropic case (or to 
a single privileged observers in the autocentric case) one might go so far 
as to envisage the attribution of non-zero weighting even to situations 
where no material observer is present at all. Such an unreasonably 
overextended weighting (as exemplified by the kind of ubiquity principle 
that was implicit in Dirac's original argument in favour of his now 
discredited theory~\cite{Dicke60,Carter83} of varying  gravitational 
coupling) might make logical sense if one could imagine perceiving oneself 
to be some sort of disembodied spirit, but (as Dirac's example shows) does 
not deserve trust for scientific purposes.

\bigskip
{\bf Illustration}
\medskip

As a toy example to illustrate the application of this micro-anthropic 
principle, consider a gedanken experiment in which Schroedinger's cat, C, 
has equal chance of being awake or dreaming, as also does its master, M, 
who, if awake can see whether the cat is too, but if asleep has equal 
chances of dreaming that the cat is awake or asleep, whether or not it 
actually is. The cat is unconcerned about its master, and so has only two 
relevant mind states $\evr_1$, $\evr_2$ (awake or dreaming) with entropy 
${\cal S}={\rm log}\, 2=1$. Schroedinger has four relevant mind states, 
$\evr_3$, $\evr_4$, $\evr_5$, $\evr_6$ with ${\cal S}={\rm log}\, 4=2$, so 
his net probability is $2/3$ while the cat's is $1/3$. The conditional  
``{\color{vert}gross}'' probability $\Pr$, and absolute ``net'' probability 
$P$ for the relevant eventualities are tabulated as follows.
 
\bigskip
\noindent

%\begin{table}
%\caption{Schroedinger's Cat 
%\label{tab:cat}}
%\begin{center}
\begin{tabular}{c c c c }   
%\multicolumn{2}{c}
\hfill\vline &{$\evr_1$: C awake} \hfill\vline &{$\evr_2$: C asleep}
\hfill\vline & {M: {\color{vert}gross}$\rightarrow$net} \hfill\\ 
\hline{ $\evr_3$: M awake, sees C awake} \hfill\vline & 
$\Pr_{31}={\color{vert}1/4}$  \hfill  & $\Pr_{32}={\color{vert}0}$ 
\hfill\vline & {\color{vert}1/4} $\rightarrow$ 1/6 \hfill \\
{ $\evr_4$: M awake, sees C asleep}\hfill\vline  & 
$\Pr_{41}={\color{vert}0}$  \hfill  & $\Pr_{42}={\color{vert}1/4}$ 
\hfill\vline  &  {\color{vert}1/4} $\rightarrow$ 1/6 \hfill\\
{ $\evr_5$: M dreams C awake}  \hfill\vline & 
$\Pr_{51}={\color{vert}1/8}$ \hfill & $\Pr_{53}= {\color{vert}1/8}$ 
\hfill\vline  &  {\color{vert}1/4} $\rightarrow$ 1/6 \hfill\\
{ $\evr_6$: M dreams C asleep} \hfill\vline  & 
$\Pr_{61}={\color{vert}1/8}$ \hfill & $\Pr_{62}={\color{vert} 1/8}$ 
\hfill\vline  &  {\color{vert}1/4} $\rightarrow$ 1/6 \hfill\\
\hline \hfill{ C: {\color{vert}gross} $\rightarrow$ net} \hfill\vline  & 
{\color{vert}1/2}$ \rightarrow$ 1/6 \hfill\vline & 
{\color{vert}1/2}$ \rightarrow$ 1/6 \hfill\vline  &   \hfill \\ 
\end{tabular}

\vskip 1 cm

{\bf 13. Local Application}
\medskip

Whereas the term {\it entropic} principle has the advantage of avoiding any 
risk of misunderstanding that the range of applicability of the ansatz  
(\ref{72}) extends beyond observers of narrowly anthropic type, on the 
other hand the alternative term {\it micro-anthropic} principle has the 
advantage of advertising the applicability of the principle (as of its 
{\it proper} predecessor) not just to the entire life of an observer but to 
particular parts thereof. The question of whether one is more likely to 
find oneself to be nearer the beginning or nearer the end of one's life was 
raised in an epilogue by Leslie~\cite{Leslie96} who suggested, on the basis 
of Everett's own (confusing) presentation of his doctrine~\cite{Everett57}, 
that its continual multiplication of the number $\calN_\evr$ of relevant 
branches entailed a probability distribution that would be heavily biased 
towards the last moments of life, on the understanding that the dogma that 
all the branch channels are equally ``real'' implies that the corresponding 
anthropic probability factor should be given by ${\calP}_\evr \propto 
\calN_\evr$ (rather than by an expression of the entropic form 
${\calP}_\evr \propto {\rm log}\{ \calN_\evr\}$ that has been advocated 
here). Having safely survived, and thereby invalidated, this alarming 
prediction, Leslie arrived at the observational conclusion that 
-- as was argued on purely theoretical grounds at the beginning of this 
essay -- this particular interpretation of the Everett doctrine is untenable.  

According to the present analysis, the correct answer to Leslie's question 
is as follows. To start it is necessary to reject not only  Everett's claim 
that the relevant branches are ``real'' (which might be interpreted as 
meaning ${\calP}_\evr\propto \calN_\evr$) but also his attribution to them, 
nevertheless, of an ordinary non anthropic quantum probability weighting 
(which might be interpreted as implying the choice of a constant value for 
${\calP}_\evr$). This contradiction between Everett's preaching and his 
practice is resolved in the present approach by what is interpretable as a 
compromise, according to which the appropriate formula has the logarithmic 
form  ${\calP}_\evr\propto {\rm log}\{ \calN_\evr\}$. 
 
The replacement of a linear by a logarithmic dependence law merely 
moderates, but does not avoid, the unrealistic implication that the 
probability distribution would strongly disfavour the earlier stages of a 
lifetime if Everett's  branching metaphor were to be taken literally. It is 
therefore obvious that this aspect of what is commonly understood to be 
meant by the Everett's interpretation is also misleading, and, as remarked 
above, it is very easy to see why. The idea of a rapidly increasing number 
of relevant branch channels is something that may make sense in the case 
when, for example, one has just taken delivery of a new computer with 
entirely empty memory banks, but it will soon ceases to be valid when 
saturation sets in so that erasion becomes necessary, so as to release to 
release occupied space by a process whereby the relevant information is 
converted into Landauer entropy~\cite{Vedral}. Concerning the human case, 
parents and primary school teachers know that even small children do a lot 
of forgetting as well as learning, while as adulthood progresses the ratio 
of what is learned to what is forgotten goes on decreasing, so that it may 
ultimately become quite small compared with unity as senility sets in. This 
means that the relevant number $\calN_\evr$ of Everett branch channels 
should normally reach a maximum -- not a peak but a plateau -- in mid life. 
It is to be understood that this statement refers to a smoothed average 
over diurnal variations, because  the number of channels involved in 
conscious perception presumably undergoes considerable reduction during 
sleep, particularly during deep dreamless phases.

For practical probabilistic purposes it is only relative values that matter. 
The intrinsically interesting question of the absolute hight of the plateau 
is beyond the scope of the present investigation, but it is evident from 
physical considerations that $\calN_\evr$ it can not be nearly as large as 
the (admittedly gigantic) value of ${\rm exp}\{Q_\evr\}$, where $Q_\evr$ is 
the Dyson entropy number discussed above, which exceeds the corresponding 
Landenauer entropy $S_\evr= {\rm log}\{\calN_\evr\}$ (representing the 
amount of useful information processed~\cite{Vedral} during the perception) 
by an enormous thermodynamical waste factor $W_\evr=Q_\evr/S_\evr\gg 1$. 
(In the days before valves were replaced by transistors, the relevant waste 
factors for computers were far worse even than those of their biological 
analogues, but the spectacular progress of engineering techniques in recent 
years has brought about an amazing rate of improvement.)

 It is a noteworthy coincidence that Dyson's evaluation~\cite{Dyson79} of 
$Q_\evr$ in the human case gave a value of the same order as the Avrogadro 
number, which  is interpretable as the number of molecules in a fraction of 
the order of $10^{-3}$ of the mass of a human body. If it is supposed that 
this fraction is comparable with the fraction of the molecules that are 
active in metabolic processes, then it can be deduced that the 
corresponding metabolic turnover time must have a value of the same order 
of magnitude as the mental time interval $\Delta_\evr\tau$, of the order of 
a fraction of a second, that was used by Dyson's as basis for his 
evaluation. 

This observation -- that the estimated duration $\Delta_\evr\tau$ of a 
conscious perception is roughly comparable with a timescale characterising 
metabolic processes throughout the body --  may offer a significant clue as 
to the nature of the (still largely mysterious) mental processes involved. 
One of the things that is rather clear is that the  relevant value of 
$\Delta_\evr\tau$ can undergo considerable variation --  lengthening in 
states of hibernation for example. In so far as the solution to the problem 
posed by Leslie is concerned, what is relevant is the age dependence of 
$\Delta_\evr\tau$. My impression, with which I think most people would 
agree, is that the typical duration of a moment of consciousness is 
relatively short in early childhood and that on average  -- modulo diurnal 
fluctuations through states of shallow or deep sleep -- it increases 
monotonically throughout life. According to the formula (\ref{73}), this 
means that the maximum of the anthropic probability distribution need not 
coincide with the summit of the midlife plateau where the relevant branch 
channel number $\calN_\evr$ and its logarithm $S_\evr$ is highest, but may 
actually occur at a more youthful stage.

\vfill\eject

\begin{figure}
\centering
\epsfig{figure=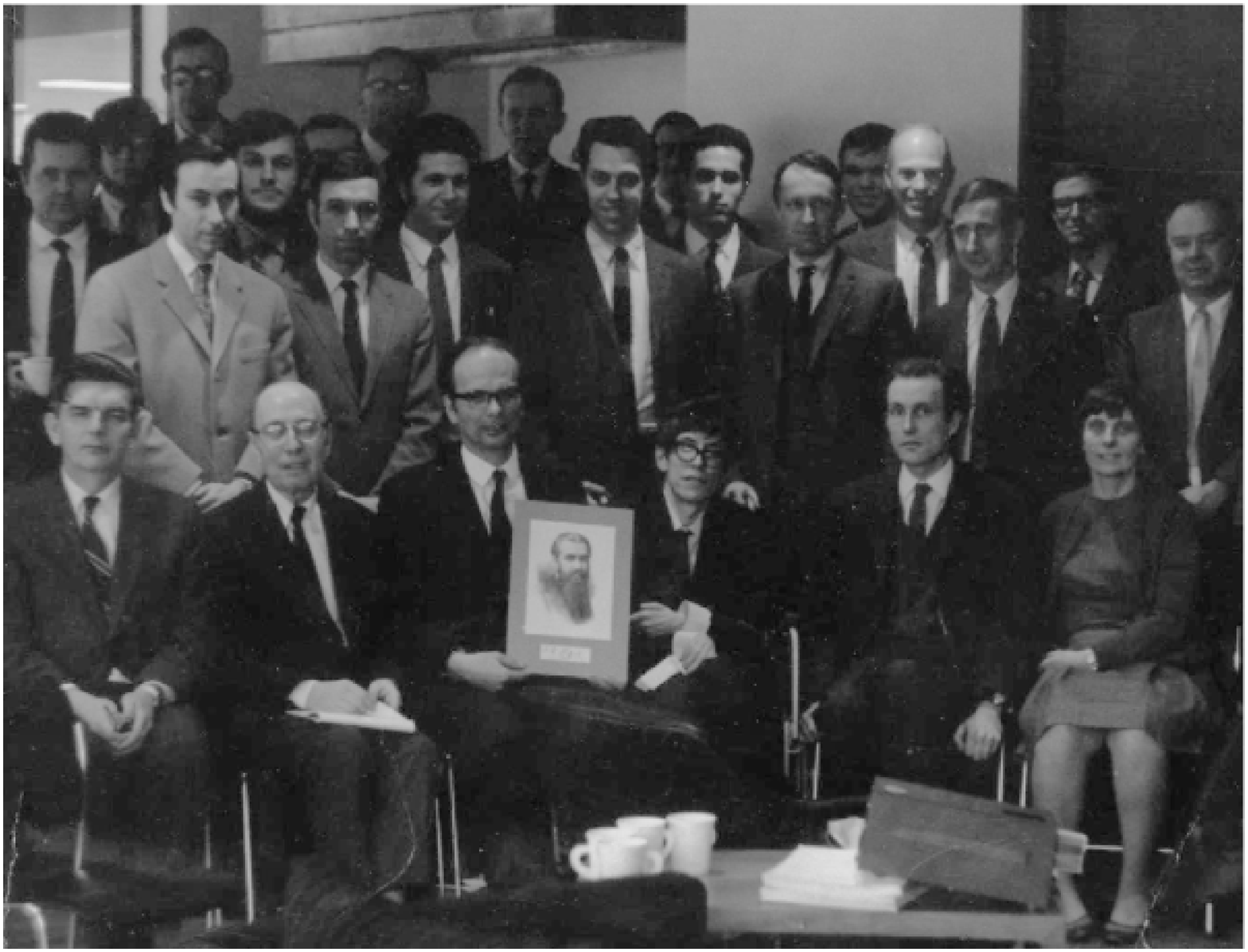, width=6in}
\caption{Some of the participants at the Clifford Centennial meeting 
organised by John Wheeler at Princeton in February, 1970, assembling many 
of the people whose thoughts contributed to the synthesis presented here, 
including, in the front at left, Bob Dicke with  Eugene Wigner, and in the 
center Stephen Hawking with the present author, behind whom are  Bryce 
DeWitt with Freeman Dyson.
 \label{fig1}}
\end{figure}


\bigskip
\begin{thebibliography}{99}

\bibitem{Carter74} B. Carter, ``Large Number Coincidences and the Anthropic 
Principle in Cosmology", in {\it Confrontations of Cosmological Theories 
with Observational Data} (I.A.U. Symposium 63) ed. M. Longair,  
(Reidel, Dordrecht, 1974) 291-298.

\bibitem{Carter83} B. Carter, ``The anthropic principle and its 
implications for biological evolution'', {\it Phil. Trans. Roy. Soc.}
{\bf A310} (1983) 347-363.

\bibitem{Carter03}  B. Carter, ``Anthropic interpretation of quantum 
theory'', {\it Int. J. Theor. Phys.} {\bf 43} (2004) 721-730.
[hep-th/0403008]

\bibitem{Everett57} H. Everett, ``Relative state formulations
of quantum theory'', {\it Rev. Mod. Phys.} {\bf 29} (1957) 454-462.

\bibitem{Graham} N. Graham, ``The measurement of relative frequency'', 
in {\it The many worlds interpretation of quantum
mechanics}, ed. B.S. De Witt, N. Graham (Princeton U.P., 1973) 229-253.

\bibitem{Wigner70} E.P. Wigner,``On hidden variables and quantum 
mechanical probabilities'',
{\it American Journal of Physics} {\it bf 38} (1970) 1005-1009.

\bibitem{Dirac} P.A.M. Dirac, {\it Quantum mechanics}
(Oxford U.P., 1958)

\bibitem{Hawking} S.W. Hawking, ``The unpredictability of quantum gravity'',
{\it Com. Math. Phys.} {\bf 87} (1982) 395-415.

\bibitem{Vilenkin} A. Vilenkin,
``Quantum cosmology and eternal inflation'',
{\it Proc. Hawking Workshop, Cambridge, 2002}
[gr-qc/0204061]

\bibitem{Zurek81} W.H. Zurek, ``Pointer basis of quantum apparatus: into
what mixture does the wave function collapse?'', 
{\it Phys. Rev.} {\bf D24} (1981) 1516-1525. 

\bibitem{Polkinghorne} J.C. Polkinghorne,
{\it One world: the interaction of science and theology},
(London, 1986).

\bibitem{Neumann} J. von Neumann, {\it Mathematical foundations of
quantum mechanics}, (Princeton U.P., 1955). 

\bibitem{Bub} J. Bub, {\it Interpreting the quantum world},
(Cambridge U.P. 1997).

\bibitem{Vedral} V. Vedral,``Landauer's erasure, error correction, and
entanglement'', {\it Proc. R. Soc. Lond.} {\bf A456} (2000) 969-984.

\bibitem{Leslie83} J. Leslie, ``Cosmology, probability, and the
need to explain life'', in {\it Scientific Understanding},
ed. N. Rescher (University Press of America, Lanham and London, 1983).

\bibitem{Carter93} B. Carter,``The anthropic principle and the 
ultra-Darwinian synthesis'', in {\it The Anthropic Principle},
ed F. Bertola , U. Curi (Cambridge U.P., 1993) 33-63.

\bibitem{DeWitt} B.S. DeWitt, ``The Many universes interpretation of 
quantum mechanics'', in {\it The many worlds interpretation of quantum
mechanics}, ed. B.S. De Witt, N. Graham (Princeton U.P., 1973).

\bibitem{Deutsch} D. Deutsch, ``Quantum theory of probability and
decisions'', {\it Proc. R. Soc., Lond.} {\bf A455} (1999)
3129-3137. [quant-ph/9906015]

\bibitem{Wallace} D. Wallace,``Quantum probability and decision
theory revisited'', {\it Studies in History and Philosophy of
modern physics} {\bf 34}(1) (2003) 87-105 [quant-ph/0107144]

\bibitem{Greaves} H. Greaves, ``Understanding Deutch's probability in
a deterministic universe''. [quant-ph/0312136]

\bibitem{Hartle} J. Hartle, ``Quantum mechanics of individual systems'',
{\it Am. J. Phys.} {\bf  36} (1968) 704-712.

\bibitem{Donald} M.J. Donald, ``A mathematical characterisation of
the physical structure of observers'', {\it Foundations of Physics}
{\bf 22} (1995) 1111-1172.

\bibitem{Lockwood} M. Lockwood, `` `Many minds' interpretations of
quantum mechanics'', {\it British Journal for the philosophy of
Science}  {\bf 47} (1996) 159-188.

\bibitem{Page} D. Page ``Sensible quantum mechanics: are probabilities
only in the mind?'', 
{\it Int. J. Mod. Phys.} {\bf D5} (1996) 583-596. [gr-qc/9507024]

\bibitem{Gell} M. Gell-Man, J.B. Hartle, ``Quantum mechanics in the
light of quantum cosmology'', in {\it Complexity, Entropy, and the
Physics of Information}, ed. W.H. Zurek (Addison Wesley, Redwood City,
1991) 425-458.

\bibitem{Bostrom} N. Bostrom, {\it Anthropic Bias: Observation
selection effects in Science and Philosophy} (Routlege, New York, 2002).

\bibitem{Leslie96}  J. Leslie,  {\it The end of the world}
(Routlege, New York, 1996).

\bibitem{Islam77} J. Islam, ``Possible ultimate fate of the universe'',
{\it Q. J. Roy. Astr. Soc.} {\bf 18} (1977) 3-8.

\bibitem{Dyson79}  F.J. Dyson, ``Time without end: physics and biology
in an open system'', {\it Rev. Mod. Phys.} {\bf 51} (1979) 447-460.

\bibitem{Dyson96}  F.J. Dyson, ``Reality bites'', {\it Nature}
{\bf 380} (1996) 296.

\bibitem{Dicke60} R.H. Dicke, ``Dirac's cosmology and Mach's principle'', 
{\it Nature} {\bf 192} (1960) 440-441.


\end{thebibliography}
\end{document}